\documentclass[journal]{IEEEtran}  % standard double-columm
\pdfoutput=1

\usepackage{graphicx} % Required for inserting images
\usepackage{amsmath,amssymb,bm}
\usepackage{color}
\usepackage{subfig}
\usepackage{url}
\usepackage{enumitem}

\title{Blind Equalization using a Variational Autoencoder with Second Order Volterra Channel Model}
\author{S\o ren F\o ns Nielsen, Darko Zibar and Mikkel N. Schmidt
\thanks{This work was supported by VILLUM FONDEN with grants MARBLE (VIL40555) and VI-POPCOM (VIL54486)}%
\thanks{S. F. Nielsen and M. N. Schmidt are with the Department of Applied Mathematics and Computer Science, Technical University of Denmark mail: sfvn@dtu.dk.}% <-this % stops a space
\thanks{Darko Zibar is with the Department of Electrical and Photonics Engineering, Technical University of Denmark}% <-this % stops a space
\thanks{Manuscript received XXXX; revised XXXX.}}

\newcommand{\E}{\mathbb{E}}
\newcommand{\EQ}{\mathbb{E}_Q}

\begin{document}
% The paper headers
\markboth{Journal of \LaTeX\ Class Files,~Vol.~14, No.~8, August~2015}%
{}%

\maketitle

\begin{abstract}
% Motivation
Existing communication hardware is being exerted to its limits to accommodate for the ever increasing internet usage globally. This leads to non-linear distortion in the communication link that requires non-linear equalization techniques to operate the link at a reasonable bit error rate.
% What are we doing about it?
This paper addresses the challenge of blind non-linear equalization using a variational autoencoder (VAE) with a second-order Volterra channel model. The VAE framework's costfunction, the evidence lower bound (ELBO), is derived for real-valued constellations and can be evaluated analytically without resorting to sampling techniques.
% Results
We demonstrate the effectiveness of our approach through simulations on a synthetic Wiener-Hammerstein channel and a simulated intensity modulated direct detection (IM/DD) optical link. The results show significant improvements in equalization performance, compared to a VAE with linear channel assumptions, highlighting the importance of appropriate channel modeling in unsupervised VAE equalizer frameworks.
\end{abstract}

\section{Introduction}
% Motivation: why do we need equalization at all.
% Internet usage -> pushing equipment to the limits -> non-linear distortion
\IEEEPARstart{I}{n} recent years, internet usage has increased dramatically in part due to the availability of video streaming and social media. Furthermore, the recent surge in training large machine learning models has led to the construction of large scale datacenters to support fast turnaround~\cite{myttonSourcesDataCenter2022}. This means that existing communications hardware is being pushed to its limits, which in many applications leads to non-linear distortion. This could for instance be saturation effects from radio frequency power amplifiers~\cite{singyaMitigatingNLDWireless2017}, the transfer function in the light emitting diode in visible light communication~\cite{yingNonlinearDistortionMitigation2015} or non-ideal modulators, chromatic dispersion and detection in short-reach optical networks~\cite{liCharacterizationNonlinearDistortion2023}. Future communication solutions need to be able to handle non-linear distortion to a larger degree than before. 

% Basics of equalization
The process of removing distortion and noise caused by the communication channel at the receiver is commonly known as \emph{equalization}. One way to optimize the equalizer is using a sequence of apriori known symbols, a \emph{pilot} sequence (a supervised approach). For linear channel distortion and intersymbol interference (ISI) a linear adaptive filter can be used, which commonly is done either through a feed-forward filter (FFE)~\cite{proakisDigitalCommunications2008} or with a combined feed-forward and feed-back filter system, denoted a decision-feedback equalizer (DFE)~\cite{yuDecisionFeedbackEqualizerOptically2007}.
% Volterra
% Motivation for Volterra (second order): VLC LED (it has a second order model :-)  - cf. Deng et al, "Mitigating LED Nonlinearity to Enhance Visible Light Communications", 2018
However, many communication channels are subject to non-linear distortion which require a non-linear equalizer to fully compensate. Popularly, this has been tackled with a Volterra equalizer~\cite{stojanovicVolterraWienerEqualizers2017}, due to its solid theoretical foundation and stability guarantees. Another class of non-linear equalizers are the \emph{neural networks} which have also attracted some attention both during their early adoption in the 1990s~\cite{GIBSON1990573}\cite{youNonlinearBlindEqualization1998} and also more recently~\cite{estaranArtificialNeuralNetworks2016}\cite{huangPerformance2022}.
% FIXME: More wide literature search to get examples from wireless communication also?

% Motivation for going blind
However, the \emph{supervised} approach of sending pilot symbols decreases the throughput of the communication system and thus much effort has also gone into investigating \emph{blind} (unsupervised) approaches. In this scheme, only knowledge of the constellation can be utilized for optimizing the equaliser weights. This was first studied for pulse amplitude modulation (PAM) formats in~\cite{sato1975a}.
% Short mention of CMA, MMA, ...
For complex-valued modulation formats, the most widely used algorithm in this category is the constant-modulus algorithm~\cite{godardSelfRecoveringEqualizationCarrier1980}. It utilizes a criterion, based on the average modulus of the constellation, to optimize a finite impulse response (FIR) filter. To accommodate for non-constant modulus constellations, an extension to CMA was proposed called the multi modulus algorithm (MMA)~\cite{yangMultimodulusBlindEqualization2002}.

% Segway to VAE
% Regardless, of the linear or non-linear structure of the equalizer we need a relevant criterion to optimize the weights. Th
More recently, a new class of blind equalization algorithms have been proposed based on a Bayesian formulation of the problem; first in \cite{caciularuBlindChannelEqualization2018, caciularuUnsupervisedLinearNonlinear2020} for the quadrature phase shift keying (QPSK) modulation format with coded data and later extended to quadrature amplitude modulation (QAM) with probabilistic constellation shaping (PCS) in \cite{lauingerBlindEqualizationChannel2022}. Both works are based on formulating the equalization problem as a variational autoencoder (VAE)~\cite{kingmaAutoEncodingVariationalBayes2022}, which tries to approximate the true posterior distribution of transmitted symbol sequence with a simpler distribution by maximizing the evidence lower bound (ELBO). It was shown in \cite{caciularuUnsupervisedLinearNonlinear2020}, that the ELBO has an analytical expression, which can be differentiated wrt. parameters of the model, if an FIR filter is used to model the channel.

% In this work...
In this work, we extend the VAE framework from \cite{caciularuUnsupervisedLinearNonlinear2020, lauingerBlindEqualizationChannel2022} to incorporate a non-linear channel model for real-valued constellations. This is achieved by assuming a specific structure in the channel (in the VAE literature known as the \emph{decoder}), namely a second order Volterra model. Our main contribution is the derivation of the Volterra VAE (V2VAE), including an analytical expression for the ELBO, and an investigation of its merits in two simulated data studies. We first show the advantage of using the non-linear channel assumptions in a Wiener-Hammerstein system. Finally, we analyze the performance of our proposed model in a simulated intensity modulated direction detection (IM/DD) optical channel, commonly found in datacenter interconnects.

% Paper structure
The rest of the paper is structured as follows. In section~\ref{sec:methods}, we first introduce the equalization problem as a VAE with linear channel assumptions\cite{caciularuUnsupervisedLinearNonlinear2020, lauingerBlindEqualizationChannel2022}. Then we propose our innovation with a second order Volterra channel model and derive the ELBO in section~\ref{sec:methods-v2vae}. In section~\ref{sec:results}, we present the results of our two numerical simulation studies; first in section~\ref{sec:results-wh} the results from the Wiener-Hammerstein system and secondly in section~\ref{sec:results-imdd} the results on IM/DD. Finally, in section~\ref{sec:conclusion}, we summarize the paper, discuss the results and present an outlook for future research.

\section{Methods}\label{sec:methods}
\begin{figure}
    \centering
    \includegraphics[width=\linewidth]{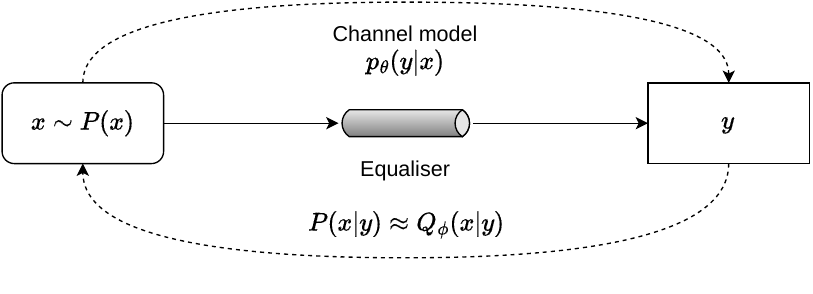}
    \caption{Variational Autoencoder (VAE) framework for blind channel equalization. In the above $\bm x$ is the (unknown) symbol sequence drawn from the prior $P(\bm x)$ and $\bm y$ is observed sequence at the receiver. The VAE then attempts to find an approximate posterior, $P(\bm x | \bm y) \approx Q_\phi(\bm x | \bm y)$, by learning the channel parameters, $\theta$, and equaliser, $\phi$, jointly.}
    \label{fig:vae-block}
\end{figure}

% Equalization
Casting the problem of channel equalization as a variational autoencoder (VAE)\cite{caciularuUnsupervisedLinearNonlinear2020, lauingerBlindEqualizationChannel2022} can be shown in the following way (see also Figure~\ref{fig:vae-block} for a visual representation). We observe a signal at the receiver, $\bm y$, from which we want to estimate the symbol sequence, $\bm x$. Given a likelihood function, $p_\theta(\bm y | \bm x)$, in which $\theta$ is the collection of all channel parameters, then the posterior can be written using Bayes rule,
\begin{equation}
    P(\bm x | \bm y) = \frac{p_\theta(\bm y | \bm x) P(\bm x)}{p(\bm y)} = \frac{p_\theta(\bm y | \bm x) P(\bm x)}{\int p_\theta(\bm y| \bm x) P(\bm x) d\bm x}\label{eq:bayes}
\end{equation}
in which $P(\bm x)$ is the symbol prior modeled as a categorical distribution, which is either uniform or adapted using probabilistic constellation shaping (PCS)~\cite{choProbabilisticConstellationShaping2019}. For all simulations in this paper, a flat prior has been used.

The posterior exactly represents what is the desired output of an equaliser, i.e. what is the most likely symbol sequence given the received signal. In most practical applications, evaluating the integral in the denominator of \eqref{eq:bayes}, the model \emph{evidence}, is practically infeasible as it involves integrating over all possible symbol sequences. Thus, we must resort to approximate methods, where the VAE comes into play. In the variational approximation, we seek a simpler distribution, $Q_{\phi}(\bm x | \bm y)$ with free parameters $\phi$, that is "close" to the true posterior, $p(\bm x | \bm y)$, in some sense. A common choice is the Kullback-Leibler (KL) divergence, which can be written as,
\begin{equation}
    \mathrm{KL}\left(Q_\phi(\bm x | \bm y)\;\|\;P(\bm x | \bm y)\right) = \int Q_\phi(\bm x | \bm y) \log \frac{Q_\phi(\bm x | \bm y)}{P(\bm x | \bm y)} d\bm x. \label{eq:kldiv}
\end{equation}

Expanding \eqref{eq:kldiv}, inserting the posterior from \eqref{eq:bayes} and rearranging the terms, can yield the following relation
\begin{align}
    \log p(\bm y) \geq &\int \log \left[p_\theta(\bm y | \bm x)\right] Q_\phi(\bm x | \bm y) d\bm x \nonumber\\
    &- \mathrm{KL}\left(Q_\phi(\bm x | \bm y)\;\|\;P(\bm x) \right), \label{eq:elbo}
\end{align}
which is known as the evidence lower bound (ELBO). Maximizing the ELBO leads to a minimization of the KL-divergence in \eqref{eq:kldiv}.

% Blockdiagram
% Good drawing -> channel + equaliser with distributions?

% Linear VAE
% Gaussian likelihood and channel model
Let the symbol sequence come from a real-valued constellation, i.e. $\bm x \in \mathcal{D}_M^N$ where $\mathcal{D}_M = \{A_1, ..., A_M\}$. Choosing a Gaussian likelihood with isotropic noise and furthermore assuming a finite impulse repsonse (FIR) filter to model the channel, we arrive at the real-valued version of the VAE derived in \cite{caciularuUnsupervisedLinearNonlinear2020, lauingerBlindEqualizationChannel2022}. We denote the time-lagged vector $\bm x_n = \left(x_n, x_{n - 1}, ..., x_{n - L} \right)$, containing all the $L$ time-points needed to produce the channel output, $y_n$. The log-likelihood function in that case can be written as,
\begin{align}
 \log p_\theta(\bm y | \bm x) = -\frac N2\log(2\pi\sigma^2) - \frac{1}{2\sigma^2} \sum_{n=1}^N (y_n - \bm x_n^T \bm h)^2, \label{eq:vae_loglike}
\end{align}
in which $\bm h$ is the FIR filter modeling the channel response and $\sigma^2$ is the noise variance, both of which are learnable parameters in the VAE framework, i.e. $\theta = \{\bm h, \sigma^2 \}$. 

We now need to specify the approximate posterior, the Q-distribution. Following a very common practice in the variational Bayesian literature, namely that the Q-distribution factorizes, we can write it as,
\begin{align}
    Q_\phi(\bm x | \bm y) &= \prod_n Q_\phi(x_n | \bm y)\label{eq:q_meanfield}
\end{align}
in which $\phi$ is the collection of all parameters of the Q-distribution and the distribution $Q_\phi(x_n | \bm y)$ is the discrete probability distribution over the constellation for the n'th symbol in the sequence. We define the individual probabilities pr. symbol through the parametric function,
\begin{align}
    f_\phi(x_n=A_m | \bm y) = Q_\phi(x_n=A_m | \bm y).
\end{align}
For more on the parameterization of the function $f_\phi(\cdot)$ cf. section~\ref{sec:meth-equalizers}.

% How to optimize? Gradient-based
The first term in the ELBO \eqref{eq:elbo}, the expectation of the log-likelihood with respect to the Q-distribution, now becomes~\cite{caciularuUnsupervisedLinearNonlinear2020, lauingerBlindEqualizationChannel2022},
\begin{align}
    \EQ \left[ \log p_{\theta}(\bm y | \bm x) \right] = &-\frac N2\log(2\pi\sigma^2) \nonumber\\
    &- \frac{1}{2\sigma^2} \underbrace{\EQ \left[\sum_{n=1}^N (y_n - \bm x_n^T \bm h)^2\right]}_{C}\label{eq:vae_lin_C}
\end{align}

Analyzing a single element of the sum inside $C$ in \eqref{eq:vae_lin_C}, denoted $c_n$, one can show that this is equivalent to,
\begin{align}
    c_n =& y_n^2 - 2 y_n \EQ[\bm x_n]^\top \bm h + \EQ[(\bm x_n^\top\bm h)^2] \nonumber\\ 
    =& y_n^2 - 2 y_n \EQ[\bm x_n]^\top \bm h + (\EQ[\bm x_n]^\top \bm h)^2 \nonumber\\
    &+ (\EQ[\bm x_n^2]-\EQ[\bm x_n]^2])^\top \bm h^2 \label{eq:vae_lin_C_analytic}
\end{align}
in which the expectations $\EQ[\bm x_n]$ and $\EQ[\bm x_n^2]$, due to the structure of the Q-distribution~\eqref{eq:q_meanfield}, are calculated independently per timepoint. A single element, $x_i$, has expectations,
\begin{align}
    \EQ\left[ x_i \right] = \sum_{m=1}^M f_\phi(x_i=A_m | \bm y) A_m \\
    \EQ\left[ x_i^2 \right] = \sum_{m=1}^M f_\phi(x_i=A_m | \bm y) A_m^2.
\end{align}

The entire ELBO is differentiable wrt. $\theta$ and $\phi$, and if we multiply the ELBO with -1, can thus be optimized using stochastic gradient descent, \emph{regardless} of the equaliser parameterization (as long as the equaliser is differentiable). To estimate the noise variance, $\sigma^2$, we use the plug-in trick from \cite{caciularuUnsupervisedLinearNonlinear2020, lauingerBlindEqualizationChannel2022}, which is achieved by analytically differentiating the ELBO wrt. to $\sigma^2$, equating to zero and solving for $\sigma^2$. This yields the solution $\sigma^2 = C / N$, which is applied in each iteration before the gradient update of the remaining paramters. Inserting the expression for $\sigma^2$ into the negative ELBO yields the loss function,
\begin{equation}
    \mathcal{L}(\theta, \phi, \bm y) = \mathrm{KL}\left(Q_\phi(\bm x | \bm y)\;\|\;P(\bm x) \right) + N\log C\label{eq:vae_loss}
\end{equation}

The KL-divergence term in \eqref{eq:vae_loss} can be calculated as,
\begin{align}
\mathrm{KL}&\left(Q_\phi(\bm x | \bm y)\;\|\;P(\bm x) \right) = \nonumber \\
    &\sum_{n=1}^N \sum_{m=1}^M f_\phi(x_n=A_m | \bm y) \log \frac{f_\phi(x_n=A_m | \bm y)}{P(x_n = A_m)}
\end{align}
which in the case of a flat prior over symbols simplifies to the negative entropy of the Q-distribution. In all numerical simulations we have used a flat prior.

% Note on non-linear channel model
It should be noted that the VAE framework technically can be used with an arbitrary non-linear channel model. However, in that case, we can no longer calculate the gradient of the loss function analytically and must resort to approximate methods. It was shown in \cite{caciularuUnsupervisedLinearNonlinear2020}, that this can be done utilizing the Gumbel-Softmax approximation~\cite{jangCategoricalReparameterizationGumbelsoftmax2017a}. This approach involves sampling the gradients, which we would expect yields more noise in optimization and has not been explored in this paper.

% Mention sps > 1?
Commonly, equalization is performed in an oversampled domain with multiple samples per symbol (sps) leading to $\bm y$ and $\EQ[\bm x]$ not having equal length. In this case the expectation vectors are upsampled by inserting zeros between the symbols to match the length of $\bm y$ as suggested in~\cite{lauingerBlindEqualizationChannel2022}.

\subsection{Variational Autoencoder with Second Order Volterra Channel Model}\label{sec:methods-v2vae}
We now turn to the case, where the channel model is assumed to have a second order Volterra series structure, and we derive an analytical expression for the expected log-likelihood. Given the time-lagged vector $\bm x_n$, then the second order Volterra model can be written as
\begin{equation}
    \hat{y}_n = \bm x_n^\top \bm h + \bm x_n^\top \bm H \bm x_n,
\end{equation}
in which $\bm h$ and $\bm H$ are the first and second order Volterra kernels, respectively. The matrix $\bm H$ is symmetric, i.e. $H_{ij} = H_{ji}$. We have for simplicity assumed here that the both the first and second order Volterra terms use the same input vector $\bm x_n$.

As in \eqref{eq:vae_loglike} for the first order Volterra model, we are interested in deriving the analytical expression for the expected log-likelihood of the VAE, more specifically the subterm $C$. Now, replacing the first order model with a second order one, we arrive at
\begin{equation}
    C = \EQ\left[ \sum_{n=1}^N \left( y_n - \left( \bm x_n^\top \bm h + \bm x_n^\top \bm H \bm x_n \right) \right)^2 \right] \label{eq:volvo_C}
\end{equation}
Analyzing a single element of the sum in \eqref{eq:volvo_C}, we arrive at
\begin{align}
    c_n =& \EQ\left[ (y_n - \hat{y}_n)^2 \right] \nonumber\\
          =& \EQ\left[ \left(y_n - (\bm x_n^\top \bm h + \bm x_n^\top \bm H \bm x_n) \right)^2 \right] \nonumber\\
          =& y_n^2 - 2 y_n \EQ[\bm x_n]^\top \bm h - 2 y_n \EQ[\bm x_n^\top \bm H \bm x_n] \nonumber\\
           & + \EQ[(\bm x_n^\top\bm h)^2] + 2\EQ[\bm x_n^\top\bm h \bm x_n^\top \bm H \bm x_n] \nonumber\\
           & + \EQ[(\bm x_n^\top \bm H \bm x_n)^2]\label{eq:volvo_c_long}
\end{align}

The terms $\EQ[\bm x_n]^\top \bm h$ and $\EQ[(\bm x_n^\top\bm h)^2]$ are identical to terms found in \eqref{eq:vae_lin_C_analytic}. In the following we will, look at each of the last two terms and derive how they can be calculated analytically, given the moments of $x_n$ assumed to follow our $Q$ distribution. The derivation of the two terms follows the same structure, namely to write out the expectations as summations, identify the matching indices such that the expectation can be simplified ($\E[x_i x_i] = \E[x_i^2]$) and appropriately subtracting terms that arise from doing full summations. For a more detailed derivation, including all terms from \eqref{eq:volvo_c_long}, we refer the reader to the supplementary material.

In the following, we will use a simplified notation where we suppress the subscript Q in the expectation, the index $n$ is removed and indices $i,j,k$ and $\ell$ are implicitly summed over, e.g. $\EQ[\bm x_n^\top\bm h] = \E[x_i h_i]$. Using this notation we arrive at,
\begin{align}
    \EQ[\bm x_n^\top\bm h &\bm x_n^\top \bm H \bm x_n]
    = \E[x_ix_jx_kh_iH_{jk}] \nonumber\\
    =& \E[x_i]\E[x_j]\E[x_k] h_iH_{jk} \nonumber\\
        & + \left(\E[x_i^2]\E[x_j]-\E[x_i]^2\E[x_j]\right) \left(2 h_iH_{ij} + h_jH_{ii}\right) \nonumber\\
        & + \left(\E[x_i^3] - 3\E[x_i^2]\E[x_i] + 2\E[x_i]^3]\right)h_iH_{ii}
\end{align}

The squared second order kernel term becomes,
\begin{align}
    \EQ[&(\bm x_n^\top \bm H \bm x_n)^2] = \E[x_i x_j x_k x_\ell H_{ij} H_{k\ell}] \nonumber\\
    =& \E[x_i] \E[x_j] \E[x_k] \E[x_\ell] H_{ij} H_{k\ell} \nonumber\\
    & + (\E[x_i^2] \E[x_j] \E[x_k] - \E[x_i]^2 \E[x_j] \E[x_k])\;\cdot  \nonumber\\
    &\qquad (2 H_{ii} H_{jk} + 4 H_{ij} H_{ik}) \nonumber\\
    & + (\E[x_i^2] \E[x_j^2] - \E[x_i^2] \E[x_j]^2 - \E[x_i]^2 \E[x_j^2] + \E[x_i]^2 \E[x_j]^2)\;\cdot \nonumber\\
    & \qquad (2 H_{ij} H_{ij} + H_{ii} H_{jj}) \nonumber\\
    & + (\E[x_i^3] \E[x_j] - 3\E[x_i^2] \E[x_i] \E[x_j] + 2 \E[x_i]^3 \E[x_j])\;\cdot\nonumber\\
    & \qquad (4 H_{ii} H_{ij}) \nonumber\\
    & + \left( \E[x_i^4] + 12 \E[x_i^2] \E[x_i]^2 - 3 \E[x_i^2]^2 \right. \nonumber\\
    & \qquad -4\E[x_i^3]\E[x_i] - 6\E[x_i]^4) H_{ii}^2\label{eq:v2vae-elbo-sqH}
 \end{align}

 The resulting loss function has the same structure as the VAE from the previous section, cf. \eqref{eq:vae_loss}, with the newly derived term $C$ \eqref{eq:volvo_C} inserted.
 
 We implemented the model and loss function using PyTorch\footnote{Our code is available from: \url{https://github.com/sfvnielsen/volterra-vae}} to allow for easy-to-use automatic differentiation and optimization routines. Our implementation follows the structure from \cite{lauingerBlindEqualizationChannel2022}\footnote{We would like to give kudos to the authors from \cite{lauingerBlindEqualizationChannel2022} for putting their code on Github.}.

\subsection{Equalizer Parameterizations}\label{sec:meth-equalizers}
Both the VAE from \cite{caciularuUnsupervisedLinearNonlinear2020, lauingerBlindEqualizationChannel2022} and the proposed extension (denoted V2VAE) support an arbitrary equaliser parameterization. To investigate the impact of the channel modeling assumptins in the likelihood, we will choose the same equalizer parameterization for both frameworks, namely a second order Volterra equalizer. However, the standard Volterra equalizer does not output probabilities and thus we apply a soft-demapping step similar to \cite{lauingerBlindEqualizationChannel2022} after the equalizer output. Assuming that the output of the Volterra equalizer is $\hat{x}_n$, then the equalizer distribution function is,
\begin{align}
    \Tilde{f}_{m,n} =& \frac{-(\hat{x}_n - A_m)^2}{\beta_m \sigma^2} \nonumber \\
    f_\phi(x_n = A_m | \bm y) =& \frac{e^{\Tilde{f}_{m,n}}}{\sum_{m'} e^{\Tilde{f}_{m',n}}}\label{eq:soft-demapping},
\end{align}
in which $\beta_m$ are noise scaling weights pr. constellation symbol, which are learned alongside all the other parameters with gradient descent. In other words, the output of the equalizer is evaluated under a (non-normalized) Gaussian density function with the constellation points as the mean and a scaled noise variance from the likelihood term. The density values are then normalized (with softmax) to yield a probability distribution. The VAE from \cite{lauingerBlindChannelagnosticEqualization2022} uses a similar soft-demapping, which for uniform symbol prior is equivalent to \eqref{eq:soft-demapping} with flat noise scaling over symbols, $\beta_m = 1, \forall m \in  \mathcal{D}_M$.

\section{Numerical Results}\label{sec:results}
\begin{figure*}
    \centering
    \includegraphics[width=\textwidth]{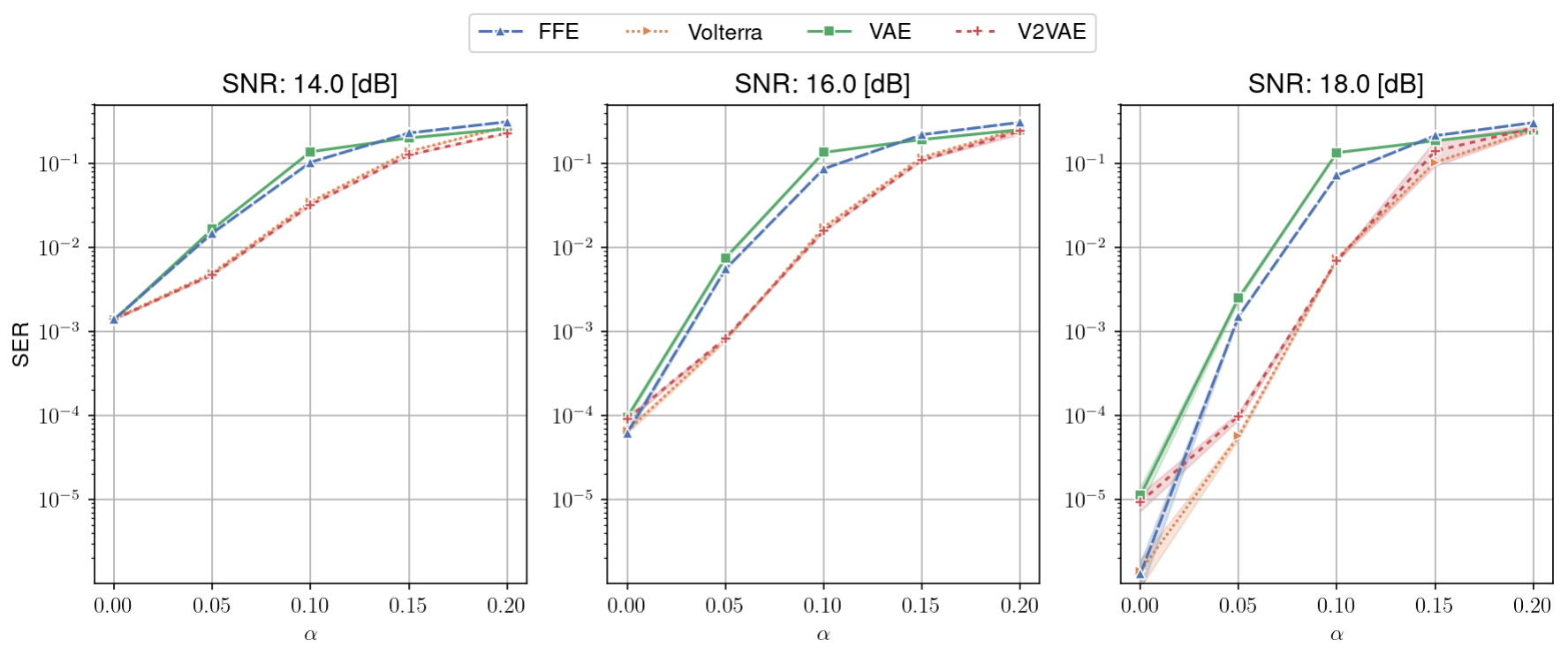}
    \caption{Symbol error-rate (SER) results for Wiener-Hammerstein channel with varying degree of non-linearity, $\alpha$. Each simulation was restarted 10 times, thus the curves represent the average over runs with shaded areas representing a 95\% confidence interval.}
    \label{fig:wh-synth-ser}
\end{figure*}

We present numerical results from two different simulation models, a Wiener-Hammerstein channel and an intensity modulated direct detection (IM/DD) optical communication system. In the results, we compare four equalization schemes,
\begin{description}[leftmargin=4\parindent,labelwidth=!]
    \item[\textbf{VAE}] Variational autoencoder from \cite{caciularuUnsupervisedLinearNonlinear2020,lauingerBlindEqualizationChannel2022} with a second order Volterra equaliser.
    \item[\textbf{V2VAE}] Variational autoencoder with second order Volterra channel model (this work) and a second order Volterra equaliser.
    \item[\textbf{Volterra}] Standard second order Volterra series equaliser with pilots (non-blind).
    \item[\textbf{FFE}] Standard linear feed-forward equaliser with pilots (non-blind).
\end{description}

We stress that both the VAE and the V2VAE are capable of doing (blind) non-linear equalization, but it is in their channel assumptions that they differ.
In all simulations, we used the Adam optimizer~\cite{kingmaAdamMethodStochastic2015} and screened the learning rate between the values $[5\cdot 10^{-3}, 5\cdot 10^{-4}, 5\cdot 10^{-5}]$. In the following, the reported symbol error rate (SER) values are for the best performing learning rate. We used a step-wise learning rate scheduling, where, given a number of iterations, $N_\text{iter}$, the learning rate was reduced every $N_\text{iter}/10$ iteration, such that the final learning rate was 10 times lower than the initial value. We use $10^6$ symbols for training and $10^6$ symbols for SER calculation after convergence.
In all simulations, we used $N_{\text{taps}}^{(1)} = 25$ taps in the FIR part of the equalizer (all methods), $N_{\text{taps}}^{(2)} = 15$ taps in the second order equalizer kernel (VAE, V2VAE and Volterra) and a channel memory $N_{\text{channel}} = 25$ (VAE and V2VAE).

\subsection{Wiener-Hammerstein Channel}\label{sec:results-wh}
The Wiener-Hammerstein system is a well-studied general function~\cite{billingsIdentificationSystemsContaining1982}, used to model non-linear dynamic systems such as loudspeakers in acoustic echo-cancellation systems~\cite{halimehNeuralNetworkBasedNonlinear2019}, power amplifiers in radio communication~\cite{bolstadIdentificationCompensationWienerHammerstein2011} and transmitters in optical communication~\cite{sasaiWienerHammersteinModelIts2020}, to mention a few. The Wiener-Hammerstein system is comprised of two finite impulse response (FIR) filters, with a memory-less non-linearity, $g(\cdot)$, in-between. We choose $g$ to be a second order polynomial and the system transfer function, $h_\text{wh}(x)$, can be written as,
\begin{align}
    h_\text{wh}(x) &= h_2 * \left(g\left( h_1 * x \right)\right)\label{eq:wh-channel},\\
    g(x) &= (1 - \alpha) x + \alpha x^2\nonumber.
\end{align}

The Wiener-Hammerstein system is inserted into a simple additive white Gaussian noise (AWGN) communication channel. We generate a random sequence of symbols drawn from the constellation $\mathcal{A} = \left\{ -3, -1, 1, 3\right\}$, also known as pulse-amplitude modulation with order 4 (PAM-4). The symbols are then up-sampled by an oversampling factor of $4$ after which they are pulse-shaped with a root-raised cosine (RRC) filter with rolloff $\rho = 0.1$. The signal is then passed to the Wiener-Hammerstein system from~\eqref{eq:wh-channel}. The FIR filters have coefficients, $h_1 = [1.0, 0.3, 0.1]$ and $h_2 = [1.0, -0.2, 0.02]$ (designed to each be minimum-phase and have two zeros) and the non-linearity $\alpha$ is varied during the simulations. Both $h_1$ and $h_2$ are upsampled (zero-insertion) with the same oversampling factor as the symbols before they are applied in the channel. After the Wiener-Hammerstein system, AWGN is added to yield a pre-specified signal-to-noise ratio (SNR) of $\frac{\E_s[h_\text{wh}(x)^2]}{\sigma^2}$, where $\E_s$ is the empirical average energy-per-symbol and $\sigma^2$ is the noise variance. Matched filtering (RRC) is applied, the sequence is decimated to 2 samples per symbol and the resulting signal is synchronized with the symbol sequence. The different equalizers are then fitted to the resulting sequence. After convergence, a new test set is generated with the same steps as above and the symbol error rate (SER) is calculated. For all methods we used a batch size of $1000$. For the two supervised methods (FFE and Volterra), we map the output of the equalizer to the nearest constellation point in the standard Euclidean sense and count the errors. For the two variational auto-encoders (VAE and V2VAE), we use the estimated symbol probabilities from the Q-distribution, pick the most likely symbol under the model and then count the errors.

The results of varying the SNR and the non-linearity coefficient $\alpha$ can be seen  in Figure~\ref{fig:wh-synth-ser}.
In the linear regime ($\alpha = 0$), we see that all methods generally achieve similar SER, i.e. the non-linear methods can adapt their non-linear components to the problem at hand, with an exception in the high SNR case where non-linear methods incur a small penalty. As we increase the quadratic term ($\alpha > 0$), we see that, unsurprisingly, the SER of the linear FFE starts to increase more than than the supervised Volterra method. The \emph{unsupervised} V2VAE closely follows the performance of its supervised counterpart, whereas the VAE with mismatched channel assumptions, follows more the trend of the linear FFE.

\begin{figure*}
    \centering
    \includegraphics[width=\textwidth]{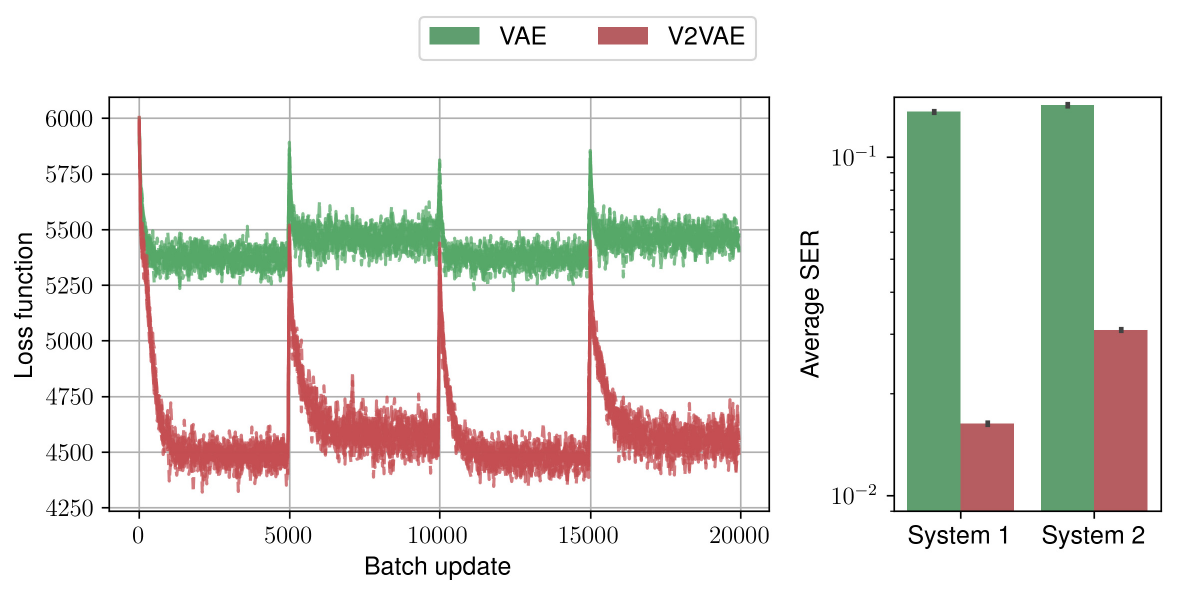}
    \caption{Tracking in Wiener-Hammerstein channel ($\alpha = 0.1$ and SNR = 16 dB). We changed the coefficients of the first FIR filter in the system every $2.5\cdot10^6$ symbols, alternating between two sets of coefficients, denoted System 1 and System 2. The test was restarted 10 times with a new seed. We show the loss function (left panel), for all restarts, as a function of batch update (batch size of 500 symbols) and the average SER (right panel) pr. system with errorbars indicating confidence interval estimated over seeds. The SER was calculated on an independent validation set of size $10^6$ symbols.}
    \label{fig:wh-tracking}
\end{figure*}

% Tracking experiment
To investigate the unsupervised methods (VAE and V2VAE) convergence properties and ability to track changes in the system, we devised a simple change-point test. We use the Wiener-Hammerstein system described above with $\alpha = 0.1$, but change the first FIR filter, $h_1$, to a new set of coefficients, $h_1^* = [1.0, 0.5, 0.1525]$, after $2.5 \cdot 10^6$ symbols and continue to alternate between $h_1$ and $h_1^*$ every $2.5\cdot 10^6$ symbols. During this simulation we disable the learning rate scheduling and run with a fixed learning rate. We report the loss function value as a function of processed batches and the average SER on a held-out validation set ($10^6$ symbols) over systems (System 1 with $h_1$ and System 2 with $h_1^*$) in Figure~\ref{fig:wh-tracking}.

In general, the V2VAE converges to a lower loss function value compared to the VAE in this non-linear channel, where the VAEs channel model assumptions is violated. As expected, the lower loss function value also translates to better SER performance, seen in the right panel of Figure~\ref{fig:wh-tracking}. However, we also note that the V2VAE converges slower to a stable loss-function level than the VAE. We attributes this to the more complex channel model that the V2VAE has to fit. 

\subsection{Intensity Modulated Direct Detection System}\label{sec:results-imdd}
\begin{figure*}
    \centering
    \subfloat[Voltage-to-optical characteristic of the modulator.]{\includegraphics[width=0.49\textwidth]{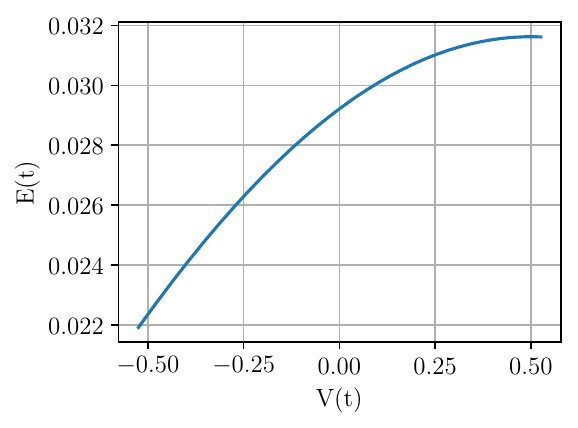}}\hfill
    \subfloat[Eyediagram after matched filter (noiseless case).]{\includegraphics[width=0.49\textwidth]{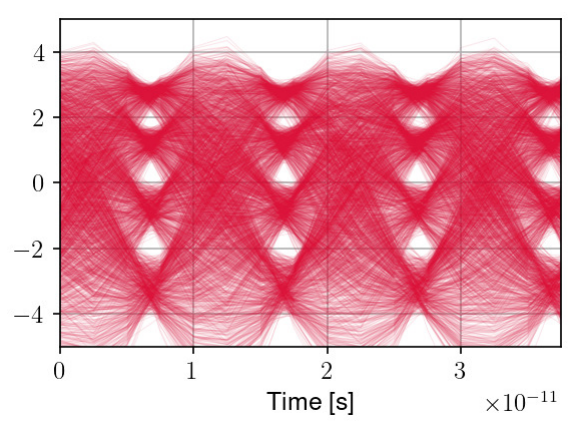}}
    \caption{Mach-Zehnder modulator (MZM) and a accompanying eyediagram after the receiver filter for $V_{pp} = 1.2$, $V_{\pi} = 2$ and $V_b = -\frac{1}{2}$. No noise was added in the photodiode, such that the eyediagram only shows the impact of the non-linearity.}\label{fig:mzm}
\end{figure*}

% IM/DD with EAM/MZM and Photodetector (100 GBaud)
In this section, we study a simulated optical communication system based on intensity modulated direct detection (IM/DD), commonly found in datacenter interconnects~\cite{zhongDigitalSignalProcessing2018}, inspired by the system model in~\cite{liangGeometricShapingDistortionLimited2023}. We again use the PAM-4 modulation format with constellation $\mathcal{A} = \left\{ -3, -1, 1, 3\right\}$. We use the same transmitter processing as in section~\ref{sec:results-wh} (up-sampling by 4 and pulse-shaping with RRC and rolloff $\rho=0.1$) with a baud rate $R_s = 100$ GBaud. The signal is then passed to a digital-to-analog converter (DAC), comprising of a voltage normalization step to the range $[-\frac{1}{2}, \frac{1}{2}]$, multiplication with a peak-to-peak voltage, $V_{pp}$ and application of a 5th order Bessel low-pass filter. The 3-db cutoff frequency of the low-pass filter was set to $55$ GHz. The voltage signal, $V(t)$, is then used as input to a Mach-Zehnder modulator (MZM), which yields the optical signal,
\begin{equation}
    E(t) = \sqrt{P_{in}} \cos\left( \frac{1}{2 V_{\pi}} (V(t) + V_b)\right),\label{eq:mzm}
\end{equation}
in which $P_{in}$ is the power of the laser at the input of the modulator and $V_{\pi}$ and $V_b$ are parameters of the MZM. A plot of the modulator characteristic and eyediagrams in the noiseless case can be seen in Figure~\ref{fig:mzm}.
% Fiber + photodiode
A standard single mode fiber model~\cite{agrawalFiberopticCommunicationSystems2002} is used to model chromatic dispersion and fiber loss. The fiber has a dispersion slope $S_0 = 0.092\;\mathrm{ps/(nm^2\; km)}$, zero-dispersion wavelength $\lambda_0 = 1310\;\mathrm{nm}$, attenuation $\alpha_{smf} = 0.2\;\mathrm{dB/km}$. We use a laser wavelength of $\lambda = 1270\;\mathrm{nm}$, which yields a dispersion parameter of $D = \frac{S_0}{4}\left(\lambda - \frac{\lambda_0^4}{\lambda^3}\right) \approx -15.43\;\mathrm{ps/(nm\cdot km)}$.
At the receiver, the signal is converted to voltage domain using a square-law detector with thermal and shot noise modeled as AWGN. The noise variances are parameterized as,
\begin{align}
    \sigma_t^2 &= \frac{4\cdot k \cdot T \cdot F_s}{B\cdot Z} \\
    \sigma_s^2 &= \frac{2 e_c \left( R\cdot \mathbb{E}[|y|^2] + I_d\right) F_s}{B},
\end{align}
in which $k$ is the Boltzmann constant, $T = 293$ is the temperature of the photodiode in Kelvin, $F_s$ is the sampling frequency, $B = 55$ GHz is the assumed bandwidth of the photodiode, $Z = 50 \Omega$ is the impedance load, $e_c$ is the electron charge, $R = 1$ [A/W] is the responsivity of the photodiode, $\mathbb{E}[|y|^2]$ is the empirical average power received and $I_d = 1 \cdot 10^{-8}$ A is the dark current.
The signal is converted to digital domain again using an analog-to-digital converter (ADC) with the same bandwidth limitation and filters as the DAC. The matched filter (RRC) is applied in the digital domain, and finally the signal is down-sampled to 2 samples-per-symbol and the symbol sequence is synchronized to the received signal. The equalizers are then fitted to the sequence with a batch size of 1000 symbols and the SER is calculated in the same way as for the Wiener-Hammerstein system.

\begin{figure*}
    \centering
    \includegraphics[width=\textwidth]{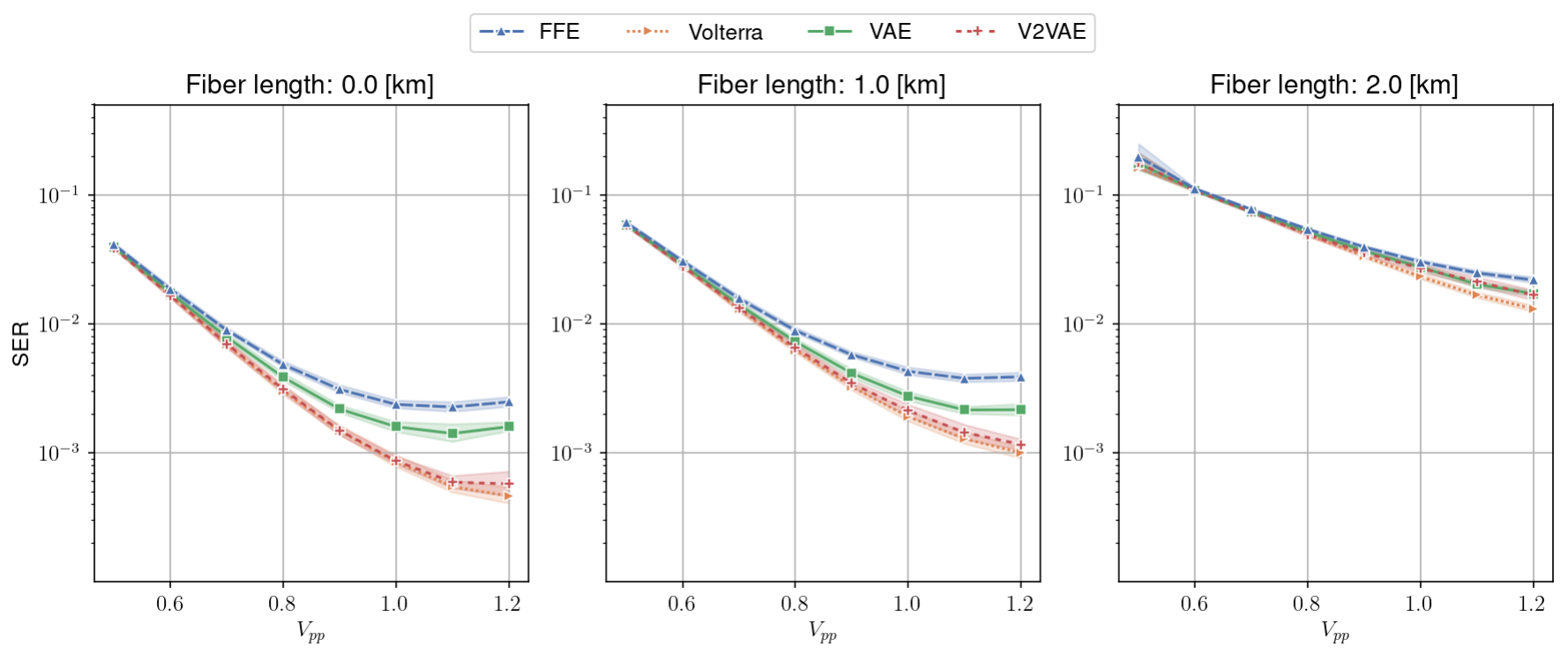}
    \caption{Results for IM/DD channel with varying fiber length and varying DAC peak-to-peak voltage, $V_{pp}$.}
    \label{fig:imdd-ser}
\end{figure*}

The SER results for varying the peak-to-peak voltage and different fiber lengths can be seen in Figure~\ref{fig:imdd-ser}. In the back-to-back (B2B) condition (fiber length of $0$ km), the main sources of distortion is the ISI introduced by the bandwidth limitation in the DAC and ADC and the non-linearity in the modulator as the $V_{pp}$ is increased. In the low voltage regime ($V_{pp} < 0.8$), there is generally little to no difference in performance across the methods, as the modulator is operating in the linear range. As $V_{pp}$ increases the non-linear methods start to gain an advantage over the supervised linear FFE. The supervised Volterra method is generally performing best, closely followed by the unsupervised V2VAE. When $V_{pp}$ reaches the highly non-linear regime, we see the biggest advantage of having a non-linear channel assumption showcased by the V2VAE performing significantly better than the standard VAE, even though they have the same equaliser parameterization. Looking at longer fiber lengths (1 and 2 km), the chromatic dispersion becomes the main source of distortion and the advantage of using non-linear methods lessens.

% Mention EAM results
We also ran the simulations with a electro-absorption modulator (EAM)~\cite{gallantCharacterizationDynamicAbsorption2008}, the result of which can be seen in the supplementary material.

\section{Discussion and Conclusion}\label{sec:conclusion}
We investigated the impact of channel modeling assumptions in a VAE equalizer framework. We extended the decoder to be non-linear with a specific structure, namely a second order Volterra series and derived the analytical ELBO for optimizing the equalizer. In both simulation studies, Wiener-Hammerstein and IM/DD, we found support that appropriate channel modeling leads to better equalization performance in the unsupervised VAE framework.
% Time-varying and tracking
% Discuss flex-update and memory efficiency
Looking at the models ability to track changes in the system, we demonstrated that V2VAE can achieve a lower cost function value compared to the VAE, but does so at a slower pace, requiring more symbols for convergence. As discussed and proposed in~\cite{lauingerBlindEqualizationChannel2022}, given enough computational resources, one could improve the convergence time by having overlapping batches and thus effectively doing more gradient updates pr. symbols (denoted the \emph{flex}-scheme in~\cite{lauingerBlindEqualizationChannel2022}). Another aspect of this is the memory efficiency, where the standard algorithms like FFE updated with LMS, can be updated once pr. symbol and only needs to store the delayline to do the gradient update. In VAE models, batched updating is preferred to reduce the variance in the stochastic gradient calculation. However, this adds a higher memory requirement in practical systems, which in turn will drive batch sizes to be as small as possible. In this paper, the minimum batch size that still yields good convergence behaviour was not explored, but could be an interesting avenue of future research.

% A note on complexity...
We note that the improvement in modeling capabilities by the V2VAE does not come for free. The calculation of the cost function and the associated gradients are more expensive to compute, compared to the linear VAE. For the VAE, the calculation of the cost function for one time-point scales with $O(L)$, as all the terms can be written as convolutions and dot products. However, the V2VAE scales with $O(L^4)$ due to the squared second order kernel term in \eqref{eq:v2vae-elbo-sqH}. We hypothesize that an efficient approximate algorithm for the V2VAE could be derived, by looking at time-averaged sufficient statistics in the ELBO, but this has not been studied yet.

% Coded data...
The authors in \cite{caciularuUnsupervisedLinearNonlinear2020} used the VAE in a low-density parity-check (LDPC) coded data transmission scenario, where the estimated symbol probabilities were used in conjunction with a belief propagation algorithm to decode the most probable bit sequence. The V2VAE also admits to this extension, and could potentially lead to better decoding performance in non-linear channels due to the more flexible channel modeling.

% Expanding to complex-valued constellations
% Coherent-detection and MIMO (cross-talk between modes)
A natural extension, would be to derive the model for complex-valued constellations, as done originally in both \cite{caciularuUnsupervisedLinearNonlinear2020} and \cite{lauingerBlindEqualizationChannel2022}. This would allow the V2VAE to be applied to coherent optical transmission and model cross-talk between different modes in the fiber as in~\cite{lauingerBlindEqualizationChannel2022}. However, that derivation has been deemed out of scope for this paper.

% Use-cases outside communication?
% Look in spiking neural network literature?
%{\color{blue}Mention potential use-cases outside equalization. Spiking data? Something where "signal-of-interest" is binary or Poission and we observe some noisy continuous version of the signal due to the system transfer function. Spiking neurons? Hemodynamic response function?}

% Future work
We used a second order Volterra model as the channel model, both due to the Volterra models popluarity for non-linear system identification and non-linear equalization. However, future work could explore other structures for non-linear channel modeling while keeping the ELBO analytical, that might computationally scale better in the channel memory.

\bibliography{TCCN_2024}

\end{document}